\documentclass[]{aastex701}

\makeatletter\def\Hy@Warning#1{}\makeatother

\begin{document}

\title{The role of environment in triggering AGN -- evidence for a change at $z\sim$1}

\author{Jason R. Reeves}
\affiliation{Department of Physics and Astronomy, Tufts University, 574 Boston Avenue, Medford, MA 02155, USA}
\email[show]{jasonree702@gmail.com}

\author[0000-0002-1917-1200]{Anna Sajina}
\affiliation{Department of Physics and Astronomy, Tufts University, 
574 Boston Avenue, Medford, MA 02155, USA}
\email[show]{anna.sajina@tufts.edu}

\author{Henry Adair}
\affiliation{Drexel}
\email{henryadair719@gmail.com}

\author[0000-0003-1748-2010]{Duncan Farrah}
\affiliation{University of Hawaii, Manoa, HI, USA}
\email{dfarrah@hawaii.edu}

\author[0000-0002-3032-1783]{Mark Lacy}
\affiliation{NRAO, USA}
\email{mlacy@nrao.edu}


\begin{abstract}

What triggers AGN in some galaxies and what role does this brief period of activity play in the overall evolution of galaxies are still open questions. This paper explores whether or not the local, on scales of $\approx$1\,Mpc, galaxy density plays a role in triggering AGN when controlling for stellar mass. We consider this question as a function of redshift and AGN selection in the X-ray vs. in the IR. We use available density maps within the 4.8\,sq.deg. XMM-LSS field in the redshift range $0.1 < z < 1.6$. Our key result is that the environment may play a role in triggering IR AGN. In particular, at $z > 1.2$ the incidence of AGN increases in higher density environments, controlling for stellar mass. However, this dependence reverses at $z < 1.2$ where the incidence of IR AGN is higher in lower density environments. By contrast, among X-ray selected AGN there is no significant local density dependence. Bootstraping analysis confirms these conclusions. While these results agree with previous work on both obscured and unobscured AGN this is the first study to use a consistent methodology across IR and X-ray samples, as well as study IR dependence in this full redshift range. Upcoming large spectroscopic surveys such as the Prime Focus Spectrograph (PFS) galaxy evolution survey will be critical in further elucidating how the environment affects AGN triggering across different cosmic epochs.

\end{abstract}


\section{Introduction} \label{sec:intro}

Active Galactic Nuclei (AGN) represent the brief periods of activity for the supermassive black holes which we believe reside at the centers of all massive galaxies. Both the $M_{BH}-\sigma$ relation \citep{magorrian98} and the similarity of the cosmic SFRD and BHAR histories \citep{Madau2014} point to a co-evolution of these SMBHs and their host galaxies. Most recently, {\sl JWST} has uncovered a large population of early heavily dust obscured AGN \citep[e.g.][]{Greene2024} leading to updated theories on how this co-evolution might have taken place especially within the first billion years of cosmic evolution \citep[e.g.][]{Silk_2024}. One thing is clear, despite their transient nature, AGN are closely inter-twined with the evolution of their host galaxies \citep[e.g.][]{Wang2024_nature} and the build up of stars in the Universe in general. Therefore understanding their demographics, evolution and triggering mechanisms are all crucial outstanding questions. These studies are all complicated by the fact that any given AGN indentification method is subject to selection biases \citep{Lacy2015,Hickox2018}.  Accounting for both obscured and unobscured (as well as ideally low and high accretion rate AGN) is needed for a complete picture of the cosmic build-up of supermassive black holes \citep{Runburg2022,Ananna2022}. 

An important aspect of understanding AGN triggering and its putative role in regulating the star-formation of a galaxy is constraining whether or not the incidence of AGN depends on the local environment of galaxies. Environment of course can be defined in a variety of different ways, but here we predominantly focus on the local (order Mpc-scale) galaxy density field which is also roughly the size of a cluster-sized dark matter halo. Such a link is expected since, in the nearby Universe, the most massive black holes are associated with massive ellipticals which reside in the densest environments \citep{Dressler1980}. 
Powerful AGN across redshifts are associated with higher mass dark matter halos as seen via clustering analyses in both observational and theoretical studies \citep[see e.g.][]{He2018,Singh2023}. 
Thanks to their intrinsic high luminosity, such AGN are often used as tracers to find high density peaks at cosmic noon and earlier epochs \citep[e.g.][and references therein]{Silva2015}. Similar clustering properties have been used to support evolutionary scenarios such as powerful QSOs  being preceeded by a heavily obscured hyperluminous IR galaxy (HLIRG) phase \citep{Farrah2004}. 

However, since regardless of AGN content, more massive galaxies reside in denser environments \citep[e.g.][]{Kauffmann2004,Laigle2018}, it is important to gauge whether this environmental dependence of AGN is just a by-product of their predominance in more massive galaxies or whether there is an additional role of the environment in AGN triggering. Such a dependence might be both positive, i.e. due to processes that drive gas to the centers of galaxies such as cold gas accretion or mergers \citep[see e.g.][]{zamojski11}, or negative i.e. due to processes such as ram pressure stripping that take gas away and lead to overall galaxy quenching \citep{Peng2010,Muzzin2013_clusters}. Early results with the SDSS dataset suggest that, controlling for stellar mass, local [OIII]-selected AGN are more prevalent in lower density environments, which may be the result of their association with actively star-forming galaxies which in the local Universe are strongly associated with lower density environments \citep{Kauffmann2004}. This is consistent with the recent study of \citet{Aradhey2025} for local variability-selected AGN.  \citet{Powell2018} find the $z<0.1$ {\sl SWIFT}-BAT AGN are preferentially in group-scale environments, as opposed to cluster-scale ones, but \citet{Powell2022} do find that the black hole mass depends on the dark matter halo mass beyond its dependence on the stellar mass suggesting an additional environmental dependence. By contrast, after controlling for stellar mass, \citet{Yang2018} found no residual environmental dependence, either positive or negative, out to $z\sim3$ for X-ray selected AGN. However, similar to \citet{Powell2018}'s more local study,  \citet{Bornancini2017} find that obscured IR-selected AGN at cosmic noon tend to live in denser environments than the unobscured ones. In summary, we find a variety of studies in the literature which show contradictory results with some suggesting no environmental dependence and others showing positive dependence, where AGN are more common in higher density environments, and others a negative dependence, where AGN are more common in lower density environments. Much of these discrepancies can be attributed to differences in methodology, especially in how the AGN's environment is quantified as well as differences in AGN selection method, which often probe different SMBH populations in different stages of their evolution. Different AGN selection methods are biased towards different typical SMBH masses and accretion rates \citep[for a review see][]{Hickox2018}. 

In this paper, we address the question of whether or not two different AGN selection methods, X-ray and IR-selected, will show the same or different environmental dependence while adopting a consistent definition of their environmental density.  We use a similar methodology to that of \citet{Yang2018}, and look for density dependence in the incidence of AGN after controlling for stellar mass. However, we use a field that is $\approx 2.4\times$ larger than in the one in \citet{Yang2018} allowing for better statistics at the high luminosity end. We explore how the incidence of AGN might be affected by the local galaxy density in two redshift bins that together span $z\sim0.3-1.6$. The paper is organized as follows. In Section\,\ref{sec:data} we present our data including our X-ray and IR AGN samples as well as photometric redshifts, density maps, and stellar masses for the parent sample of galaxies in our field. In Section\,\ref{sec:results}, we present our results on the fractions of galaxies hosting AGN in different redshift, stellar mass, and density bins. In Section \ref{sec:discussion} we discuss the connections of our work to the existing literature and the implications re possible triggering mechanisms for obscured and unobscured AGN. In Section\,\ref{sec:summary} we present our summary and conclusions. Throughout, we adopt a flat $\Lambda$CDM cosmology with $\Omega_{\Lambda}=0.714$, $\Omega_{M}=0.286$ and $H_o=69.3$\,km/s/Mpc \citep{Planck2020_cosmo_parameters}. 
 
\section{Data} \label{sec:data}
\subsection{Multiwavelength photometry and photometric redshifts} \label{sec:survey}
Our study is done within 4.8\,sq.deg. of the XMM-LSS field where there is a wealth of multiwavelength data. Full details on the field coverage are given in \citet{Krefting2020}, but here we summarize the key dataset. The field has  u-band data from the CFHTLS survey \citep{Gwyn2012},  $grizy$ data from the HSC-Deep survey \citep{Aihara2018}, near-IR $JHKs$ data from the VIDEO survey \citep{Jarvis2013}, {\sl Spitzer} IRAC 3.6\,$\mu$m and 4.5\,$\mu$m data from the SERVS survey \citep{mauduit2012}, {\sl Spitzer} IRAC 5.8\,$\mu$m and 8.0\,$\mu$m and MIPS 24\,$\mu$m data from the SWIRE survey \citep{Lonsdale2003} and {\sl Herschel} SPIRE 250, 350, and 500\,$\mu$m data from the HerMES survey \citep{Oliver2012}. We use the Tractor forced photometry catalog produced by \citet{Nyland2017} supplemented by the Herschel Extragalactic Legacy Project \citep[HELP;][]{Shirley2019} for the longer IRAC channels.  We have photometric redshifts derived from this catalog photometry using EAZY (\citet{Brammer2008}). The quality of the redshifts is assessed both by comparison with spectroscopic redshifts as well as using the pair method of \citet{QuadriWilliams2010}. Both methods show $\sigma/(1+z)\approx0.03$ out to $z\approx1.6$, but a worsening of the redshifts beyond that \citep{Krefting2020}. 

Crucial for our study, this field has X-ray coverage from the XMM-SERVS survey \citep{Chen2018} which co-adds old and new XMM-Newton observations across $\sim$5.3\,sq.deg. of the XMM-LSS field with a median exposure per pixel of 46\,ks and a total of $\sim$5300 X-ray point sources identified down to limiting fluxes of $1.7 \times 10^{-15} \frac{\text{erg}}{\text{cm}^2 \text{s}}$, $1.3 \times 10^{-14} \frac{\text{erg}}{\text{cm}^2 \text{s}}$, and $6.5 \times 10^{-15} \frac{\text{erg}}{\text{cm}^2 \text{s}}$ in the 0.5 - 2 keV, 2-10 keV bands, and 0.5 - 10 keV, respectively.

\subsection{2D Density Maps} \label{sec:densitymaps}
We use the 2D density maps generated by \citet{Krefting2020} for $\sim 481,000$ galaxies in the above field selected to have $K<23$ AB.  \citet{Krefting2020} generated 2D density maps limited to the 0.1$<z<$1.6 range. This range was sliced into 28 redshift slices and a kernel density estimate with an Epanechnikov kernel was used to find the 2D galaxy densities within each slice. A simulated lightcone was used to assess the effects of photometric redshift uncertainties. These have the general effect of smearing out overdensities \citep[see][]{Krefting2020} making the strength of any dependence on density we measure a likely a lower limit. This is why we emphasize the need for large scale spectroscopic studies as follow up since there we are not subject to such density smearing. The approach in generating density maps in \citet{Krefting2020} is the same as was used to generate the density maps for the COSMOS field \citep{Darvish2016} which were used by \citet{Yang2018} to explore the environmental dependence on the incidence of X-ray AGN. 

Each galaxy within our Tractor catalog is assigned a local density value by reading it from its position in the density map for its corresponding redshift slice. We look for environmental trends by placing galaxies in three density bins: the lower $20^{\text{th}}$ percentile (``Low Density"), the $20^{\text{th}}$ - $80^{th}$ (``Medium Density"), and the upper $80^{th}$ percentile (``High Density") bins.  While in this paper we consider much wider redshift bins (e.g. $0.1<z<1.2$), we compute these percentiles within the redshift slices used for the density map calculation from \citet{Krefting2020}.  However, we tested this and found that the effect of doing this vs. simply computing the percentiles in the full redshift slices was minimal in terms of our results. 

Lastly note that since the density field evolves with redshift, a given percentile has a different meaning at higher vs. lower redshift (e.g. is associated with different dark matter halo masses as discussed in Krefting et al. 2020). However, we focus here on the relative density effects rather the dependence on absolute values of galaxy density, which is fairly standard practice in the literature \citep[see e.g.][]{Yang2018}. 

\subsection{X-ray AGN sample} \label{sec:xray_sample}
The X-ray AGN sample of this study is drawn from the XMM-SERVS survey described in Section\,\ref{sec:survey}. of \citet{Chen2018} which co-adds old and new XMM-Newton observations across $\sim$5.3\,sq.deg. of the XMM-LSS field with a median exposure per pixel of 46\,ks and a total of $\sim$5300 X-ray point sources identified down to limiting fluxes of $1.7 \times 10^{-15} \frac{\text{erg}}{\text{cm}^2 \text{s}}$, $1.3 \times 10^{-14} \frac{\text{erg}}{\text{cm}^2 \text{s}}$, and $6.5 \times 10^{-15} \frac{\text{erg}}{\text{cm}^2 \text{s}}$ in the 0.5 - 2 keV, 2-10 keV bands, and 0.5 - 10 keV, respectively. 

The hardness ratio analysis of \citet{Chen2018} suggests the vast majority of these X-ray sources are not strongly obscured, as expected given the relatively shallow X-ray data.  

Restricting this initial X-ray sample down to the field used in this study leaves 4,505 X-ray detected sources. \citet{Chen2018} do a cross-match of their X-ray sources with the SERVS {\sl Spitzer} IRAC survey \citep{mauduit2012} which provides higher accuracy `OIR' positions in their catalog.  We cross-match these OIR positions of the X-ray sources with our host galaxies sample restricted to the range $z = 0.1 - 1.6$ where we have density maps (see above). This cut leaves a sample 2,115 X-ray AGN. Of these, 564 have spectroscopic redshifts with $> 99\%$ confidence which we adopt whenever available (see \citet{Krefting2020} for more details). We double checked that the photometric and spectroscopic redshifts are consistent, certainly within the broad redshift bins considered in this work. 

\begin{figure}[h!]
    \includegraphics[width = 0.48\textwidth]{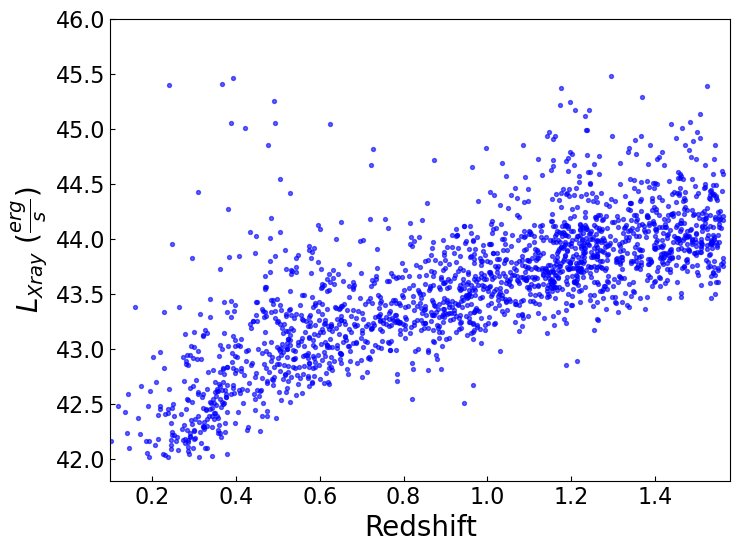}
    \includegraphics[width = 0.48\textwidth]{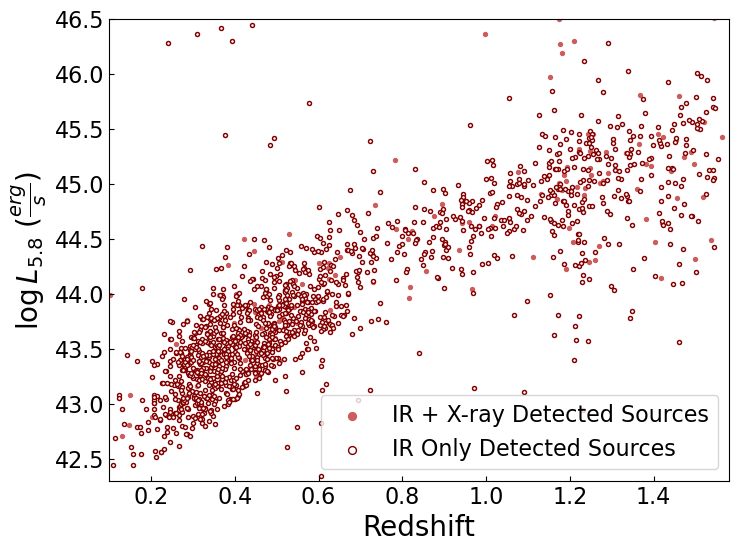}
    \caption{The redshift vs. luminosity plots for the X-ray detected AGN ({\it left}) and IR detected AGN ({\it right})  after removing sources matched to host galaxies with stellar masses $\log(\frac{M_*}{M_{\odot}}) < 9.3$ and $\log(\frac{M_*}{M_{\odot}}) > 12.0$, being outside the mass bins used in our study. Unfilled circles in the IR plot represent IR detected AGN without a matched X-ray source.
    }
    \label{fig:lum_z}
\end{figure}

Following \citet{Runburg2022}, we compute the $k$-corrected full band X-ray luminosity of our sources using:

\begin{equation}
L_x [erg/s]= 4\pi d_L(z)^2 f_x (1 + z)^{(\Gamma - 1)}
\end{equation}

where $d_L$ is the luminosity distance and $f_x$ is the observed frame 0.5$-$10\,keV X-ray flux in $\nu f_{\nu}$ from \citet{Chen2018}. The redshifts are those of the host galaxies. A spectral index of $\Gamma = 1$ is adopted following \citet{Runburg2022}. Figure\,\ref{fig:lum_z} {\it left} shows the X-ray luminosities vs. redshift for the full X-ray parent sample. An X-ray luminosity cutoff of $L_X>10^{42} \frac{\text{erg}}{\text{s}}$ is used to exclude potential non-AGN X-ray sources removing a further 5 objects from the sample \citep[see][and references therein]{Chen2018}. 

\subsection{IR AGN sample} \label{sec:ir_sample}
We select IR AGN from this dataset using the {\sl Spitzer} IRAC color cuts from \cite{lacy_obscured_2004}. Compared to other IRAC AGN selection techniques \citep[e.g.][]{stern_mid-infrared_2005, donley_identifying_2012}, the Lacy color-color cut is more complete, but suffers from greater contamination of non-AGN sources including star-forming galaxies at higher redshifts. Adding a flux cut of $S_{8\mu m}>48$\,$\mu$Jy in IRAC channel 4 decreases this contamination significantly as discussed in \citet{Berta2007}. Applying the color and flux cut selection leads to a IR AGN sample of 2,636 AGN in our field. Of these 1,832 were successively matched to host galaxies within the redshift range $0.1<z<1.6$ using a 2 arcseconds matching radius.

We calculate the rest-frame 5.8\,$\mu$m luminosities of the IR AGN using Equation\,\ref{eqn:lum58}, where $F_{5.8}$ is the rest-frame flux ($F_{5.8}\equiv \nu_{5.8} f_{\nu_{5.8}}$) $\nu_{5.8}$. To obtain the rest-frame $f_{\nu,5.8}$ from the observed IRAC photometry, we perform a quadratic fit to the observed fluxes for each source using the \texttt{numpy} module \texttt{polyfit}. 

\begin{equation}
    L_{5.8} [erg/s]=4\pi d_L(z)^2 F_{5.8}
    \label{eqn:lum58}
\end{equation}

As was done for the X-ray AGN, we use the spectroscopic redshifts when available. Figure\,\ref{fig:lum_z} {\it right} shows the 5.8\,$\mu$m luminosities vs. redshift for the full IR AGN sample. Lastly, 103 objects had issues with the quadratic fit leading to unrealistic (negative) restframe 5.8\,$\mu$m luminosities and were removed from the sample. Removing these leaves us with a final sample of 1602 IR AGN. 

We observe an overdensity of IR selected sources at $z \sim 0.4$ compared to the rest of the redshift range considered in this study. This is likely caused by contamination in the sample from galaxies with obscured star formation having high PAH emission being observed in the IRAC bands. However, very few of the objects at this redshift have high enough luminosity to be considered in the analysis of this study so their affect on the results are negligible. 

\begin{figure*}
\centering
     \includegraphics[width=0.98\textwidth]{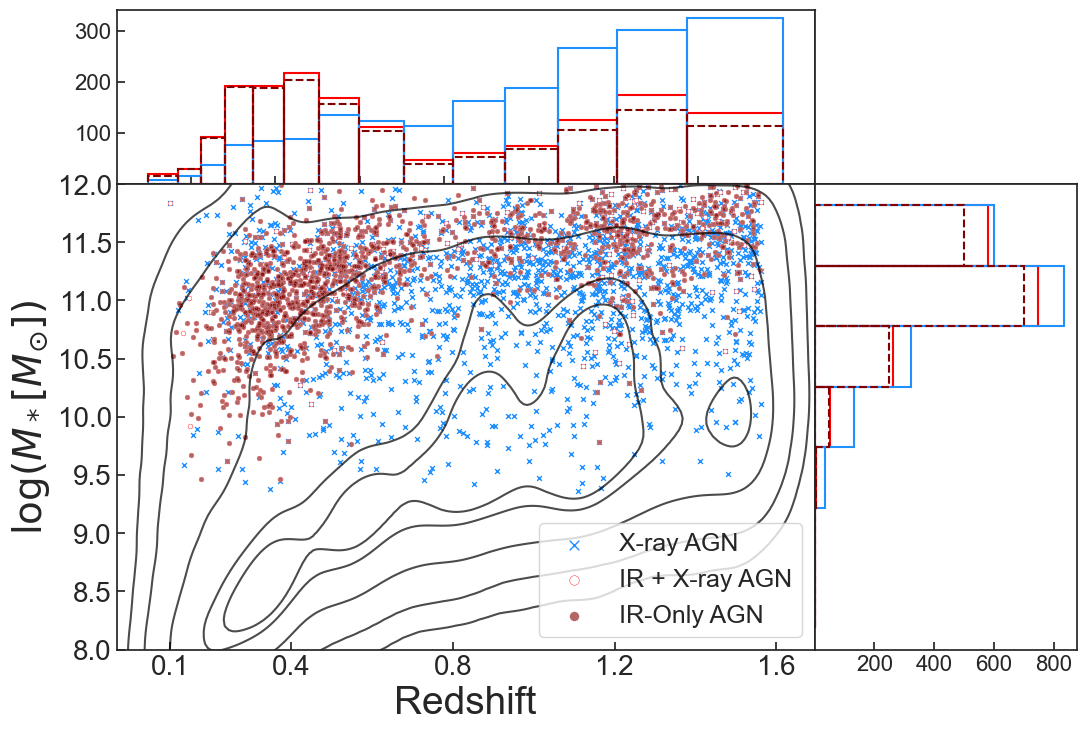}
    \caption{The contour lines show the stellar mass versus redshift for all galaxies in the parent catalog at levels of $0.1^{\text{th}}$, $1^{\text{st}}$, $10^{\text{th}}$, $50^{\text{th}}$, and $70^{\text{th}}$ percentile galaxy counts. The points represents the three different AGN samples as indicated in the legend. Histograms on the top and right hand side show the distributions of photometric redshifts and stellar masses of the host galaxies of the X-ray sources (in blue) and IR sources (in bright red), and IR-only (dashed dark red), respectively. 
    }
    \label{fig:redshift_smass}
\end{figure*}

\subsubsection{Obscured AGN Selection}\label{sec:Obscuration_Selection}

While the X-ray AGN sample can largely be considered unobscured due to their low hardness ratios, the reverse, i.e. that the IR AGN sample is consistent with obscured AGN is not necessarily true. IRAC AGN selection is largely obscuration independent, selecting both obscured and unobscured AGN \citep{Lacy2004, Hickox2018}. \citet{Lacy2015} used optical spectroscopic follow-up of mid-IR IRAC selected AGN to find that the obscured fraction among these IR AGN is largest among lower luminosity AGN and drops at the highest luminosities -- consistent for example with receeding torus models. Note that recent JWST results find that this drop in obscured AGN fraction with luminosity is not necessarily true with a more extensive IR AGN selection \cite{Lyu2024}. 

We use our X-ray sample as a proxy for the unobscured AGN population in this field. Therefore, IR sources without an X-ray counterpart should be highly obscured sources possibly even with Compton thick column densities \citep[see][]{Hickox2018}.  
We cross-matched the IR and X-ray samples using a 2\,$\arcsec$ matching radius and found 567 objects in common. Removing these from the IR sample, we are left with 1,464 obscured AGN -- from hereon our "IR-only AGN" sample. Of these, 413 have high quality spectroscopic redshifts which are adopted when available. The right-hand panel of Figure \ref{fig:lum_z} shows the sample of IR-only detected sources overlaid with the sources detected both in the IR and X-ray. We find that this obscured, IR-only, sample is roughly consistent with the full IR sample in redshift coverage. 

\subsection{Stellar masses \label{sec:masses}}

The stellar masses for all galaxies in the \citet{Krefting2020} sample are derived using the CIGALE \citep[Python Code Investigating Galaxy Emission;][]{Boquien_2019} SED-fitting code. 
In the fitting, we adopt the \citet{bz03} SSP library; the Salpeter IMF \citep{Salpeter_1955}; a delayed SFH of the form $SFR(t) = \alpha \frac{1}{\tau^2}*exp(-t/\tau)$; a dust attenuation law following \citet{CharlotFall2000}; and IR dust emission, including AGN following the \citep{Dale2014} SED templates.  The parameters that were varied were: the age of the stellar population, the stellar population model's e-folding time ($\tau$ in the range between 500\,Myrs and 12\,Gyrs considered), $A_{V,ISM}$, and a dust attenuation parameter $\mu$ which is defined as $\frac{A_{V,ISM}}{A_{V,BC}+A_{V,ISM}}$ (where BC stands for birth cloud), and lastly the AGN fraction which is the ratio of the AGN luminosity to the dust and AGN luminosity.  

The best fit model is the one with the best reduced reduced $\chi^2$ value \citep{Boquien_2019}.  CIGALE also returns the best parameters and their uncertainties which are calculated from the likelihood function exp($-\chi^2/2$) across the model grid \citep{Boquien_2019}. Note that while the specific model choices would affect the numerical values of our derived stellar masses; most of these effects are systematic across the sample (for example a different IMF or dust attenuation model would shift all masses in the same direction). This is irrelevant for the purposes of the present study where we use these stellar mass estimates only to allow us to place our AGN within stellar mass bins (allowing us to control for the stellar mass in examining environmental effects). The accuracy of our stellar mass estimates is irrelevant, as long as any systematic errors are roughly the same across the sample\footnote{We verified that this is the case by also running our analysis using a proprietary catalog in the field using newer photometry and a more sophisticated (DenseBasis) SED-fitting model. The core results of our paper remained the same.}.  

Figure\,\ref{fig:redshift_smass} shows the stellar mass vs. redshift distribution for all galaxies in the field shown as contours vs  those that host IR or X-ray AGN. As expected, the AGN hosts tend to be higher mass than the overall population, with typical host galaxy masses of $>10^{10.5}$M$_{\odot}$, although the X-ray AGN hosts in particular do have a tail toward lower masses. 

\section{Results} \label{sec:results}

\subsection{Quantifying dependence of AGN occurrence on local density} \label{sec:xray_and_ir_results}

To quantify the dependence of AGN triggering on environment while controlling for stellar mass, we use the methodology of \citet{Yang2018} where we consider the fractions of galaxies hosting AGN in different stellar mass bins and density bins. We consider four stellar mass bins ($log(\frac{M_*}{M_{\odot}}) = 9.3 - 9.975, 9.975 - 10.65, 10.65 - 11.30$, and $11.30 - 12$). We also consider three density bins (see Section\,\ref{sec:densitymaps}). Note that as discussed in Section\,\ref{sec:densitymaps} we used an adaptive density bin determination. However, we note that the effect of this was minimal and our conclusions are qualitatively the same without this step. Note also that 100 X-ray AGN, 61 IR AGN, and 24 IR-only AGN are excluded from this analysis due to their host galaxy stellar masses being outside the above range. 

The AGN percentages were calculated by dividing the number of AGN (${N_{AGN}}$) in each mass-density bin by the total number of galaxies (${N_{Gal}}$) in the same bin and multiplying the result by 100. We assume Poisson errors for the uncertainties in the number of AGN ($\sigma_{N_{AGN}}$) and total galaxies ($\sigma_{N_{Gal}}$) in each bin. If we label the occurrence rate $Q$ (i.e. $Q=100*\frac{N_{AGN}}{N_{Gal}}$) the error on that rate is given by: 
\begin{equation} 
\frac{\sigma_Q}{Q} = \sqrt{\Big(\frac{\sigma_{\rm{N_{AGN}}}}{\rm{N_{AGN}}}\Big)^2 + \Big(\frac{\sigma_{\rm{N_{Gal}}}}{\rm{N_{Gal}}}\Big)^2}
\label{eqn:quotient_error}
\end{equation}

\begin{equation}
\text{Occurrence}\% = \frac{\rm{N_{AGN}}}{\rm{N_{Galaxy}}}
\end{equation}

\begin{figure}
    \centering
        \includegraphics[scale = 0.215]{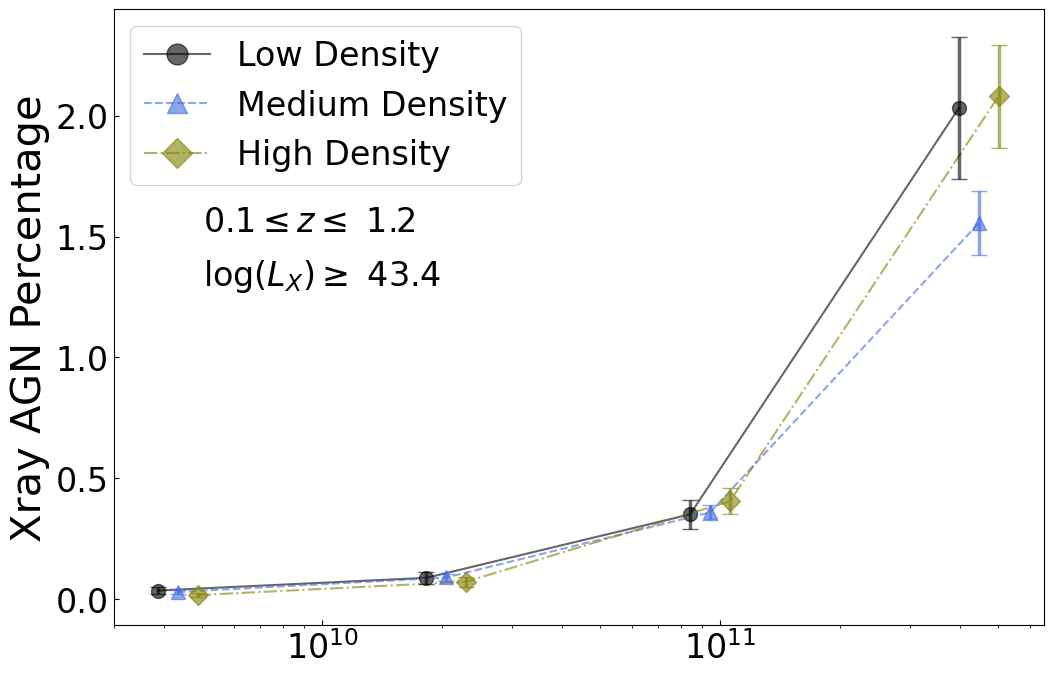}
        \includegraphics[scale = 0.215]{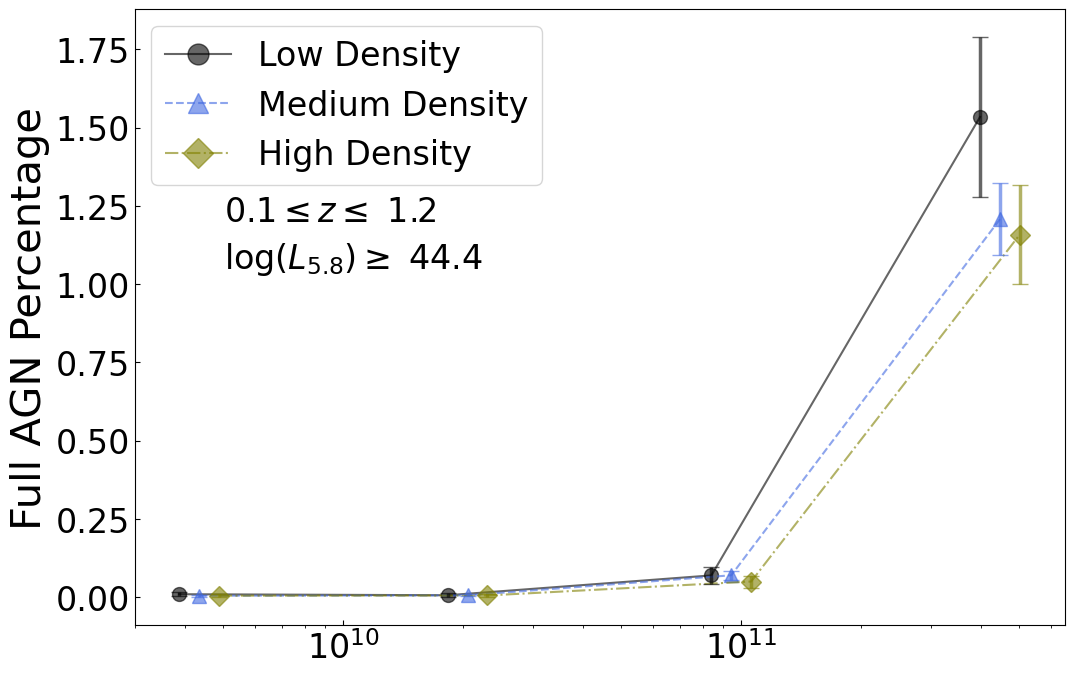}
        \includegraphics[scale = 0.215]{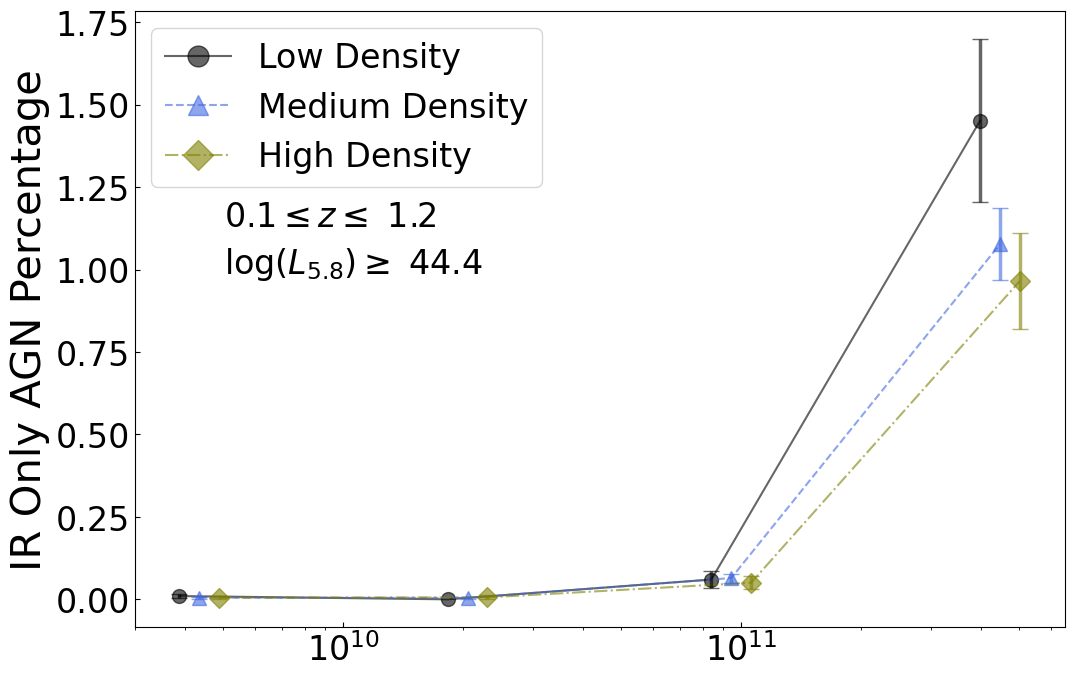}\\
        \includegraphics[scale = 0.215]{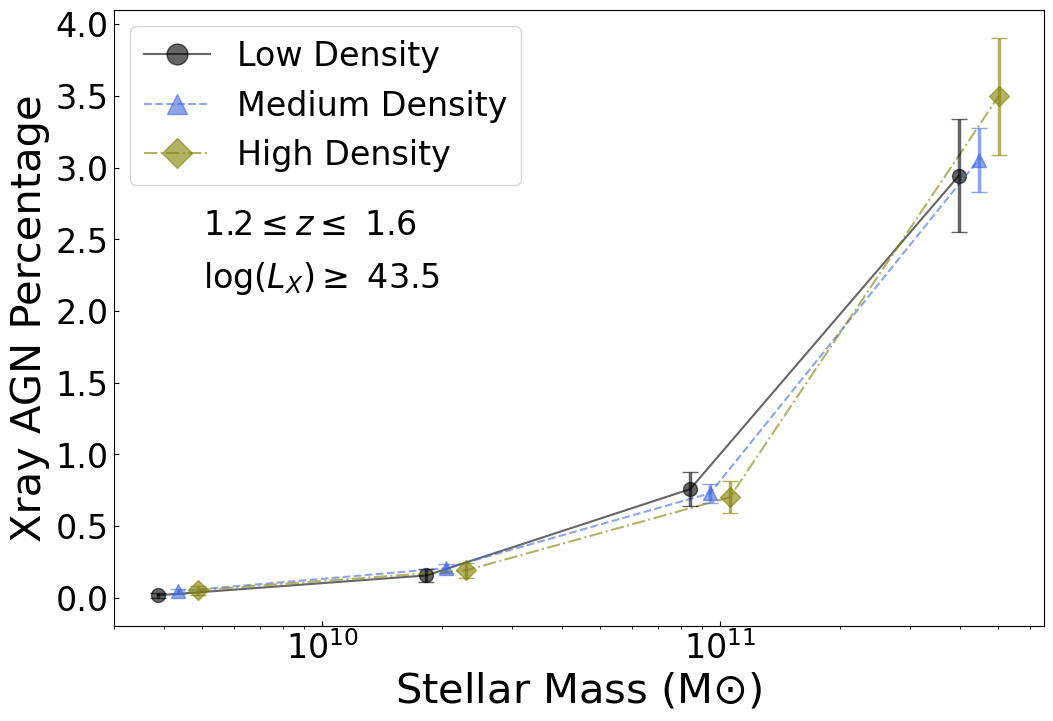}
        \includegraphics[scale = 0.215]{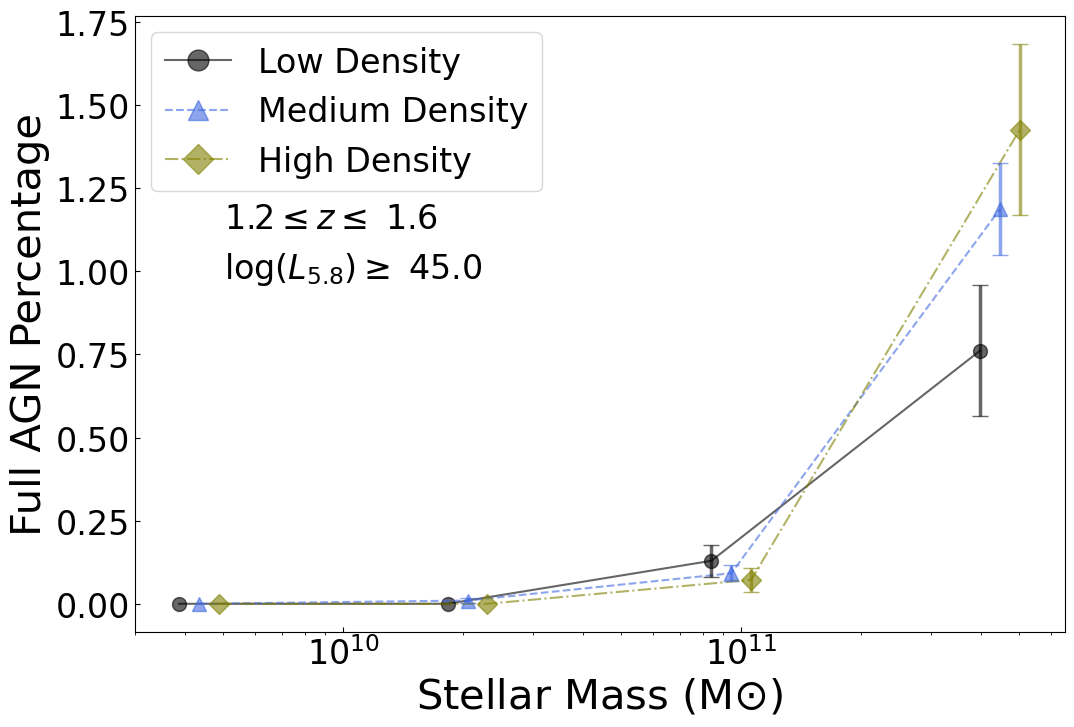}
    \includegraphics[scale = 0.215]{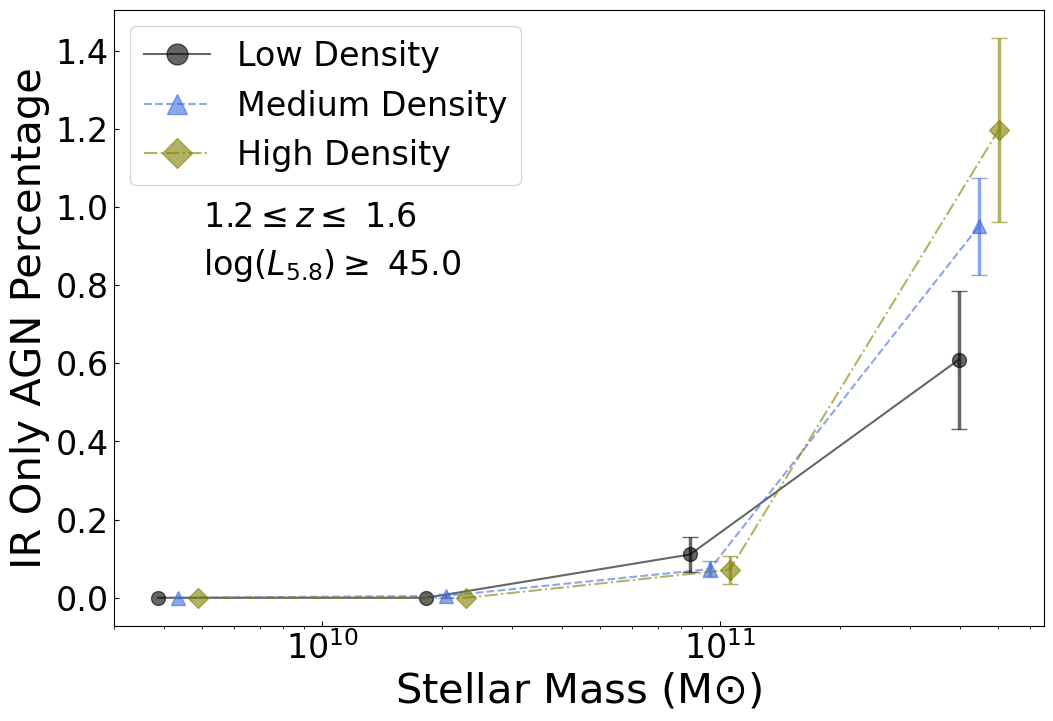}
    \caption{The AGN percentage in different stellar mass and local density bins. Here the top panels represent the lower redshift bin ($0.1<z<1.2$) whereas the bottom three panels represent the higher redshift bin ($1.2<z<1.6$). In all panels, we apply an additional luminosity cut as indicated in order to have luminosity complete samples per redshift bin. }
    \label{fig:AGN_percentages_z_l_bins}
\end{figure}

Figure\,\ref{fig:AGN_percentages_z_l_bins} shows the percentage of galaxies hosting AGN in two luminosity-complete redshift bins. 

The left-hand panels of Figure\,\ref{fig:AGN_percentages_z_l_bins} where we see no significant correlation. By contrast the middle and right-hand panels of Figure\,\ref{fig:AGN_percentages_z_l_bins} show that IR AGN experience a reversal around $z \sim 1.2$. At $z>1.2$ there is a positive correlation between AGN occurrence and environment (i.e. more AGN per stellar mass bin in higher density environments). However, IR detected AGN at lower redshifts show an negative dependence (i.e. fewer AGN per stellar mass bin in higher density environments).
 
To help us further visualize and quantify the results in the different redshift \& luminosity bins shown in Figure\,\ref{fig:AGN_percentages_z_l_bins}, we employ a scoring metric given by: 

\begin{equation}
\rm{Score} = \frac{\%_{high} - \%_{low}}{\%_{medium}}
\label{eqn:score}
\end{equation}

where $\%_{high}$ is the AGN occurrence in the highest density bin, $\%_{medium}$ is the occurrence in the medium density bin, and $\%_{low}$ is the occurrence in the lowest density bin. The metric seeks to measure the relative occurrence in the two extreme density bins while controlling for the overall amount of occurence for the sample of AGN in a bin of redshift. We apply this metric to the stellar mass bin, $log(\frac{M_*}{M_{\odot}}) = 11.3 - 12$. This metric measures which environment AGN prefer weighed by overall AGN occurrence in a certain bin of redshift and luminosity. We find scores from 0 to 0.25 to correspond to weak correlation, scores from 0.25 to 0.5 correspond to moderate correlation, and scores greater than 0.5 to correspond to strong correlation. The uncertainty of the score is again given by the quotient rule (Equation\,\ref{eqn:quotient_error}) using the uncertainties of the AGN occurrence derived earlier.

\begin{figure}
    \centering
    \includegraphics[width = 0.75\linewidth]{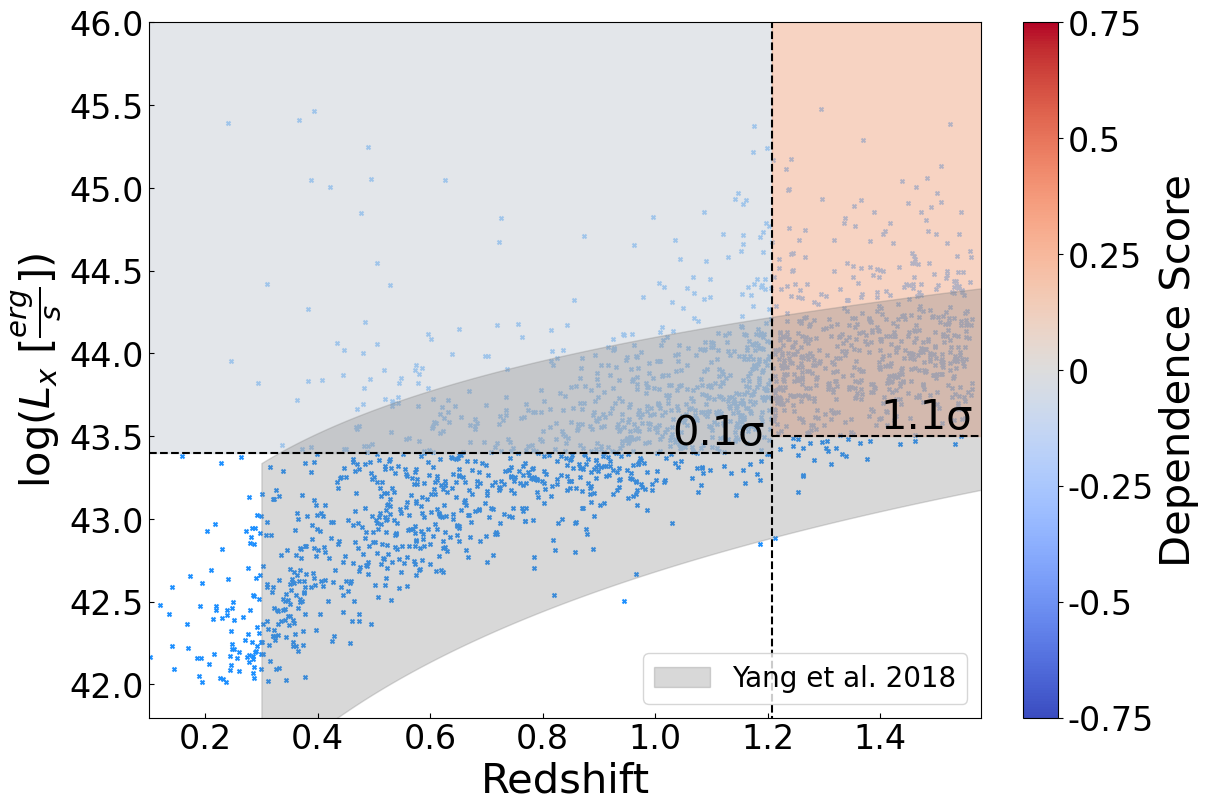}
    \caption{Redshifts versus luminosity of the X-ray sources. The gray band illustrates the X-ray AGN sample from \citet{Yang2018} who do a similar study to ours (see Section\,\ref{sec:discussion} for further discussion). The color of the bin represents the type of dependence seen in this regime of redshift and AGN luminosity. Blue bins represent regimes of anti-dependence with more dense environments having a lower percentage of galaxies hosting an AGN. Red bins represent regimes of positive correlation between the AGN occurrence and extragalactic environmental density. Gray or pale colored bins represent regimes with no dependence between the AGN occurrence and extragalactic density. The weak coloring of the bins for X-ray detected sources represents the lack of dependence seen in either bin of redshift and luminosity considered in this study. Overlaid on each bin is the significance of the score (see Section \ref{sec:xray_and_ir_results})}
    \label{fig:xray_score}
\end{figure}

\begin{figure}
    \centering
    \includegraphics[width=0.75\linewidth]{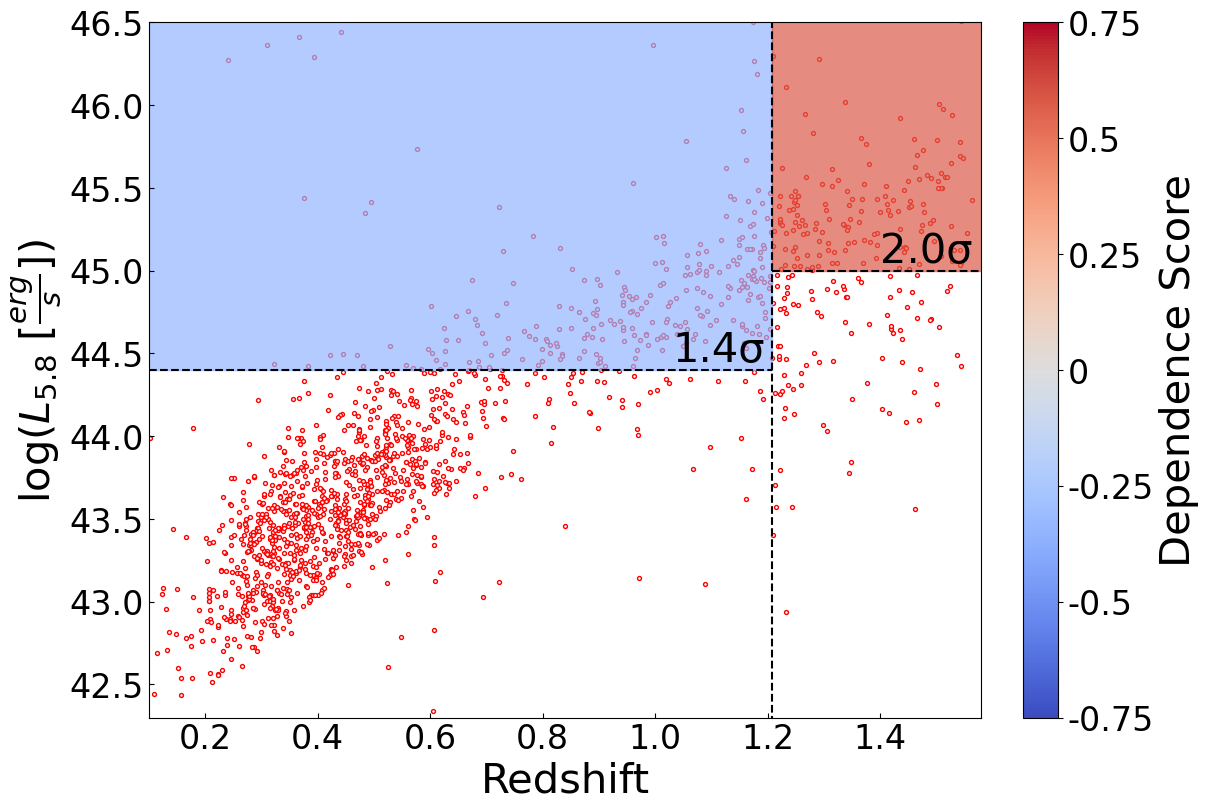}
    \caption{Redshifts versus luminosity of the full IR sample. Similar to Figure \ref{fig:xray_score}, the color of the bins correspond to the type of dependence seen in a bin of redshift and AGN luminosity with the significance of the score overlaid in each bin. Between the two bins the type of dependence exhibited for IR AGN switches at $z = 1.2$ such that it turns from a positive dependence to a negative one.}
    \label{fig:ir_score}
\end{figure}

\begin{figure}
    \centering
    \includegraphics[width = 0.75\linewidth]{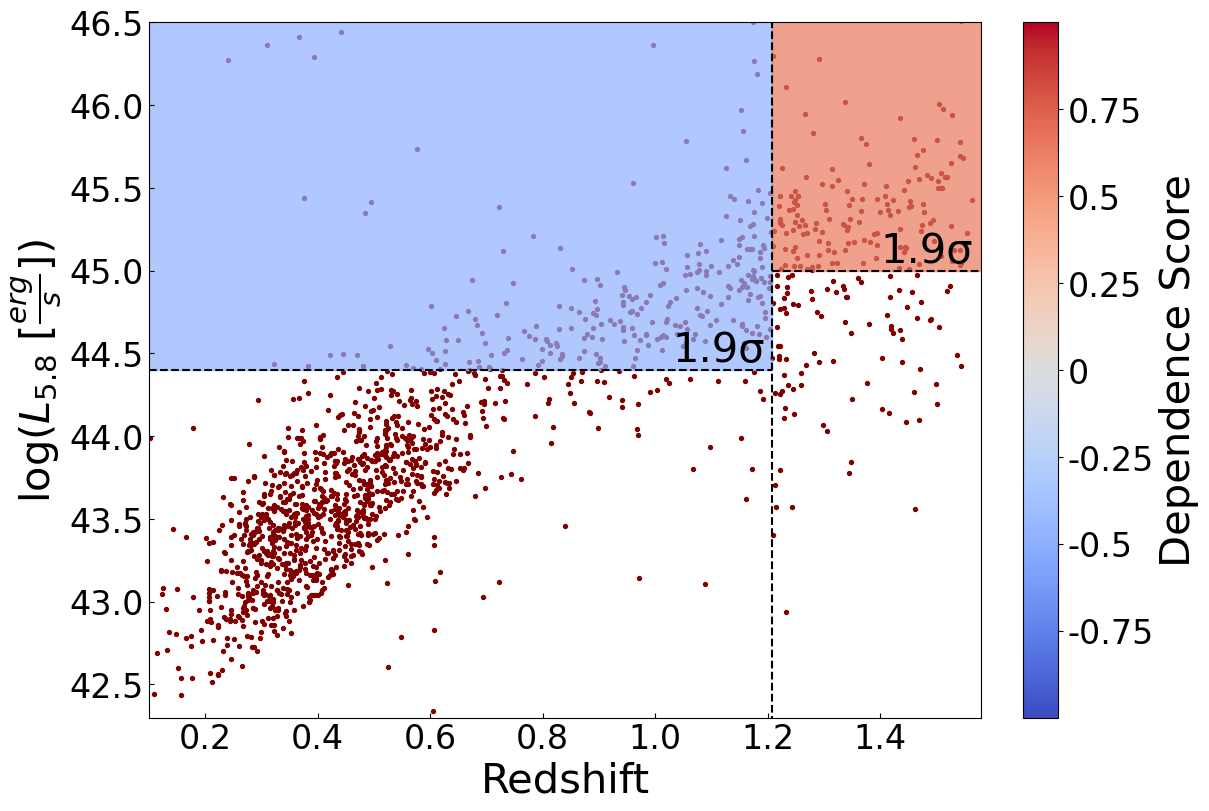}
    \caption{Redshifts versus luminosity of the IR only sample. Similar to Figures \ref{fig:xray_score} and \ref{fig:ir_score}, the color of the bins correspond to the type of dependence seen in a bin of redshift and AGN luminosity with the significance of the score overlaid in each bin. Like Figure \ref{fig:ir_score}, the two bins switch with respect to the type of dependence exhibited between the occurrence of AGN detected in the IR only and the density of their extragalactic environments. At $z = 1.2$ the dependence turns from a positive dependence to a negative one.}
    \label{fig:ir_only_score}
\end{figure}

In Figures\,\ref{fig:xray_score}, \ref{fig:ir_score}, and \ref{fig:ir_only_score} each AGN sample is split into the same two redshift bins ($0.1 < z < 1.2$ and $1.2 < z < 1.6$). Each bin is also cut in luminosity such that we have luminosity complete samples per redshift bin. The bins are color-coded by the measured score while the significance of the score (i.e. its signal-to-noise ratio) is given as the number in the bottom right corner of the panel. In addition, for Figure\,\ref{fig:xray_score} we also overlay the luminosity-redshift coverage of the \citet{Yang2018} sample (see Section\,\ref{sec:discussion} for further discussion).

Figure \ref{fig:xray_score} again shows that in the in the redshift range $0.1 < z < 1.6$ there is no significant dependence between X-ray AGN occurrence and environment. At higher redshifts there is only a weak correlation with a low confidence of $1.3\sigma$. Figures \ref{fig:ir_score} and \ref{fig:ir_only_score} show that for both samples of IR AGN, however, the $z > 1.2$ bins show positive density dependence at a confidence of $\sim 2\sigma$. Below $z\sim 1.2$, Figure\,\ref{fig:ir_score} shows moderate negative dependence for the full IR AGN sample (ie fewer AGN in denser environments) while Figure \ref{fig:ir_only_score} shows the same for the obscured IR-only sample, but with marginal significances of $1.4\sigma$ for the IR sample and $1.9\sigma$ for the IR-only sample.

\subsection{Testing the effect of modifying the specific bins}

We did multiple different iterations of this analysis to see how sensitive it is to the specific bins considered. Below we discuss our findings regarding both changes in the redshift bins as well as the luminosity bins considered. 

Regarding the redshift binning, we considered a more restricted redshift range of $0.3<z<1.2$, which we explored since the number of AGN above our completeness limits is very low in the $0.1<z<0.3$ range.  Our results were not significantly different in this scenario. 

We also explored the effects of changing the transition redshift between the two redshift bins. The behavior seen in Figures\,\ref{fig:xray_score}, \ref{fig:ir_score}, and \ref{fig:ir_only_score} remains qualitatively the same if we change the transition redshift to both lower and higher than $z=1.2$. However, the significance of the high scores in the high redshift bin is highest when $z=1.2$. Since the significance strongly depends on the number of objects per bin, this conclusion is likely not universal but a function of our specific samples. Similarly, we should not over-interpret the difference in significance between the IR samples and the IR only sample. The later is the smallest in number of AGN which explains the lower significance ($\sim2\sigma$) of the positive score in the higher redshift bin. On the other hand, the higher significance seen for the IR only sample in the lower redshift bin may be indicative of the sample of heavily obscured AGN. 

It is important that we use luminosity complete bins. Otherwise, we mix lower luminosity sources at somewhat lower redshift with higher luminosity sources at somewhat higher redshift within the same redshift bin. This would confuse our results in the presence of any luminosity-dependent effects. This is why a large number of lower luminosity AGN are excluded from the lower-$z$ bins in Figures\,\ref{fig:xray_score}-\ref{fig:ir_only_score}. We note that for the X-ray sample the luminosity complete limits are nearly the same for both redshift bins, whereas for the IR samples we have about 0.5\,dex difference. While this is not a large difference, it is conceivable that the redshift dependent effects we see for the IR AGN are partly luminosity-dependent. We attempted to explore this using multiple luminosity bins per redshift bin but found that we do not have the statistics for conclusive results.

\subsection{Testing the redshift-dependent role of environment}\label{sec:bootstrap}

\begin{figure}
    \centering
    \includegraphics[width=0.98\linewidth]{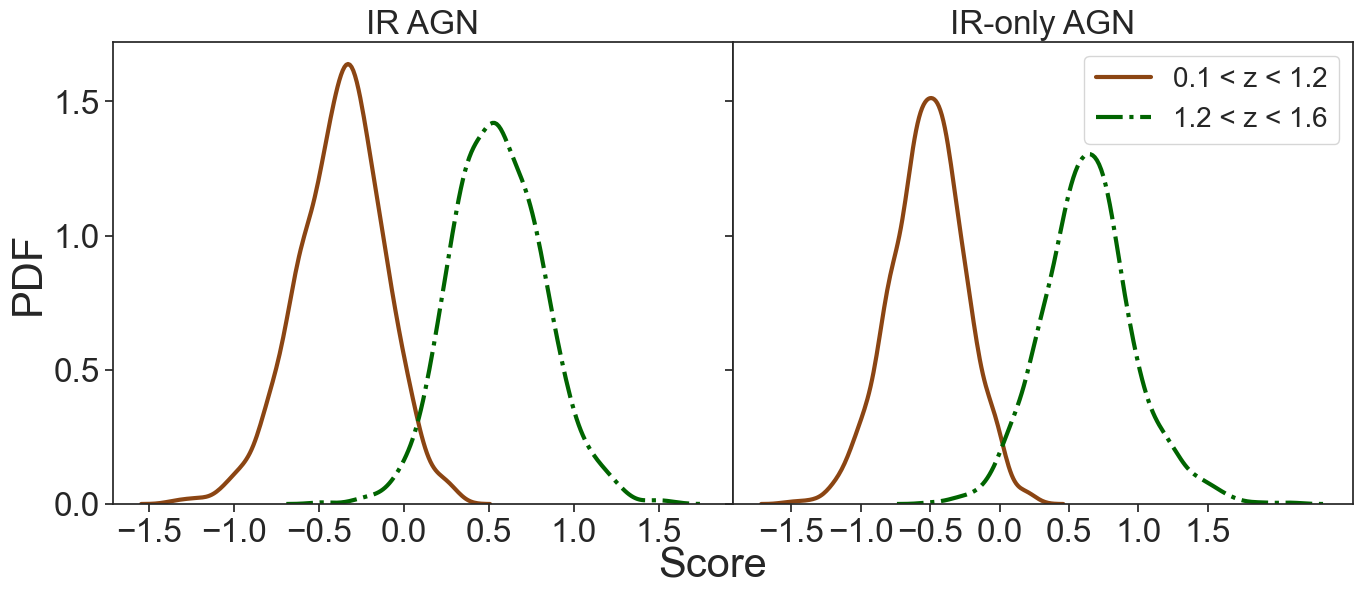}
    \caption{Probabilities of the scores from 1000 bootstrapped samples of IR and IR-only AGN in the redshift bins from Figure\,\ref{fig:ir_score} (\textit{left}) and Figure \ref{fig:ir_only_score} (\textit{right}). This analysis further strengthens our results of a reversal in the density dependence at $z\sim1.2$ with IR AGN preferring higher density environments at $z>1.2$ and lower density environments at $z<1.2$. }
    \label{fig:bootstrapped_ir_scores}
\end{figure}

Due to the low formal significance of our results in Figures\,\ref{fig:xray_score}, \ref{fig:ir_score}, and \ref{fig:ir_only_score}, we employ bootstrapping to test the strength of the results, specifically the reversal in dependence seen in both IR samples around $z\sim1.2$. We randomly resampled the IR and IR-only AGN with replacement 1000 times and calculated the scores for each using equation \ref{eqn:score}. The results are shown in Figure\,\ref{fig:bootstrapped_ir_scores}. In each IR AGN sample we see a clear difference in the peaks of the scores of the bootstrapped samples. For AGN in the redshift range $0.1 < z < 1.2$ the samples peak strongly at negative scores around $\sim -0.5$ with $95\%$ of the IR samples and $98\%$ of the IR-only bootstrapped samples having negative scores corresponding to negative dependencies between AGN occurrence and environmental density. For AGN in the redshift range $1.2 < z < 1.6$ the samples peak with positive scores of $\sim 0.5$ with $98\%$ of both the IR and IR-only bootstrapped samples having positive scores. These results highlight the result of Figures\,\ref{fig:ir_score}, and \ref{fig:ir_only_score} showing a reversal of dependence for both the IR AGN and the heavily obscured IR-only AGN samples around $z\sim1.2$. 

\begin{figure}
    \centering
    \includegraphics[width=0.5\linewidth]{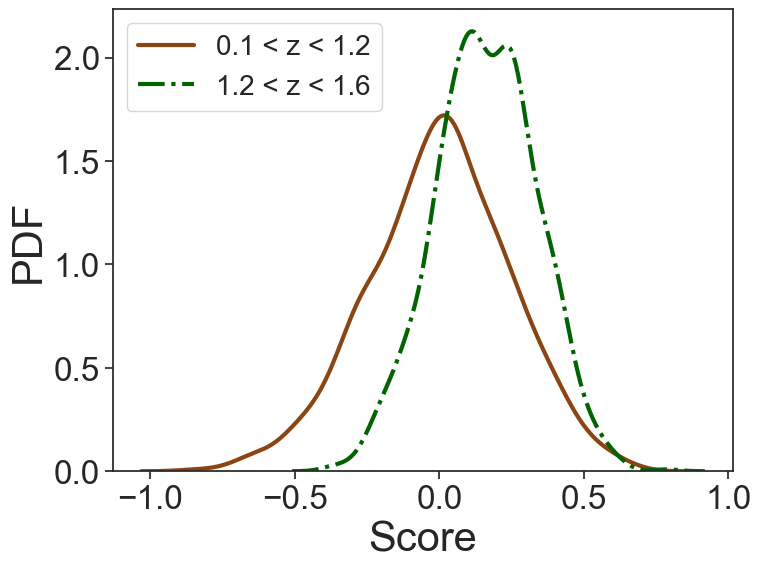}
    \caption{Probabilities of the scores from 1000 bootstrapped samples of X-ray AGN in the redshift bins from Figure\,\ref{fig:xray_score}. While there is still an offset in the peaks of the two distributions such that the higher-$z$ bootstrapped samples prefer positive scores and the lower-$z$ samples peak around zero, this offset is not statistically significant. Here 23\% of the $z>1.2$ bootstrapped subsamples show negative scores. } 
    \label{fig:bootstrapped_xray_scores}
\end{figure}

Similar analysis was carried out with the X-ray sample as shown in Figure\,\ref{fig:bootstrapped_xray_scores}. The bootstrapped X-ray samples also show positive values at $z>1.2$ whereas the distribution peaks around zero for the $z<1.2$ bootstrapped samples; however, the significance of this shift in the peak is quite small. In the $z<1.2$ bin, $56\%$ of the scores are above zero. In the $1.2 < z < 1.6$ redshift bin,  $77\%$ of the scores are above 0.

\section{Discussion} \label{sec:discussion}
\subsection{Comparison with other studies on the environmental dependence of AGN}

Our finding of no strong dependence of X-ray AGN per mass bin agrees with \citet{Yang2018}, who find no environmental dependence among a sample of X-ray detected AGN across $0.3 < z < 1.6$. Our studies differ in the depths of the X-ray data used. \citet{Yang2018} uses the \textit{Chandra} COSMOS-Legacy survey of limiting depth $8.9 \times 10^{-16} \frac{\text{erg}}{\text{cm}^2 \text{s}}$ in the 0.5 - 10 keV band \citep{Civano2016}. Our study uses the \textit{XMM-Newton} XMM-SERVS survey with limiting depth of $6.5 \times 10^{-15} \frac{\text{erg}}{\text{cm}^2 \text{s}}$ in the 0.5 - 10 keV band \citep{Chen2018}. This is a roughly 7\,$\times$ difference in depth. On the other hand, our field is $2.4\times$ larger than the one used in \citet{Yang2018}. Thus, our sample has significantly fewer lower luminosity X-ray AGN, but more higher luminosity ones, as illustrated in Figure\,\ref{fig:xray_score}. The difference in limiting depth and hence loss of lower luminosity AGN also explains why \citet{Yang2018} find 4-6\% of the highest mass galaxies host AGN vs $\approx$2-3.5\% for our sample. 

\begin{figure}[t]
    \centering
    \includegraphics[scale = 0.5]{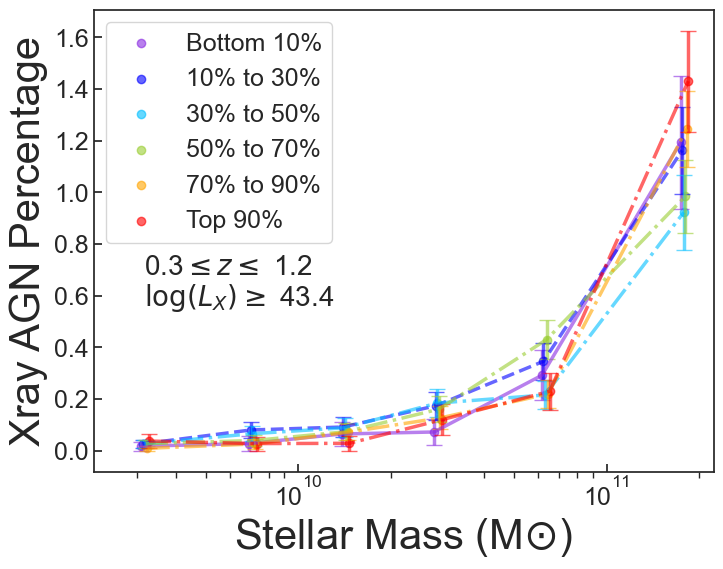}
    \includegraphics[scale = 0.5]{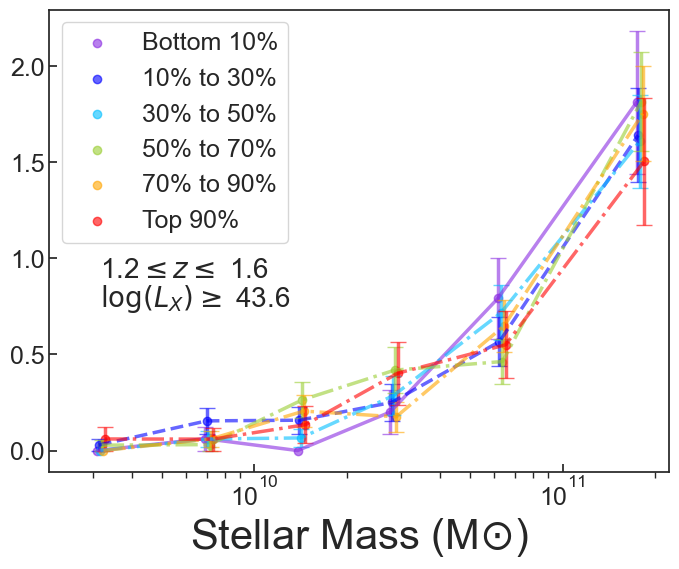}
    \caption{Recreation of Figure 10 from \citet{Yang2018} using the X-ray sample described in \ref{sec:xray_sample}, adopting their bins of redshift, stellar mass, and density. Density bins are created by splitting galaxies into percentile ranks, similar to \ref{sec:densitymaps}. Here, density bins are divided by the $10^{\text{th}}$, $30^{\text{th}}$, $50^{\text{th}}$, $70^{\text{th}}$, and $90^{\text{th}}$ percentile ranks of the extragalactic densities. In both bins of redshift we recover their result showing no significant difference in occurrence as a function of extragalactic density.}
    \label{fig:Yang_Recreation}
\end{figure} 

\citet{Yang2018} perform their analysis in three redshift bins, finding no significant density dependence in any of them. In Figure\,\ref{fig:Yang_Recreation}, we recreate their figure for the two bins where we overlap \citep[see Figure\,10;][]{Yang2018}. Note that our stellar masses assume a Salpeter IMF (see Section\,\ref{sec:masses}) whereas the ones adopted by \citet{Yang2018} use a Chabrier IMF. This results in a systematic difference of $\approx$0.25dex. For a fair comparison, in generating Figure\,\ref{fig:Yang_Recreation} we shift our sample stellar masses downward by 0.25\,dex. Our Figure\,\ref{fig:Yang_Recreation} is not a perfect recreation of \citet{Yang2018} since our sample is shallower. For example in the lower redshift bin, we are not complete down to $\log(L_X[\frac{\text{erg}}{\text{s}}])=42.6$, but rather $\log(L_X[\frac{\text{erg}}{\text{s}}])=43.4$. Within the uncertainties, we find no significant dependence on the incidence of AGN relative to the local galaxies density, consistent with the \citet{Yang2018} results. 

Among our IR samples, we do find a dependence on the environment such that at $z>1.2$ the incidence of IR and IR-only AGN increases at higher densities. This is in agreement with \citet{Bornancini2017} who find no environmental dependence for unobscured sources in the redshift range $1 < z < 2$, but a positive local density dependence for obscured sources.  However, we are able to extend our analysis to lower redshifts than that study and find that the trend reverses around $z\sim1.2$, with the incidence of IR AGN increasing at lower densities. We further verified this reversal in environmental dependence through our bootstrapping analysis. Our finding that below $z\sim1.2$ IR AGN are associated with less dense environments is reminiscent of the early SDSS result where local [OIII]-selected AGN prefer less dense environments \citep{Kauffmann2004}. Our low-$z$ results are also in agreement with the recent study of \citet{Aradhey2025} who use mid-IR variability to select AGN and find that their incidence is higher in cosmic voids relative to cosmic walls. 

Our results highlight the fact that the dependence on the incidence of AGN on their local galaxy environment differs depending on AGN selection method, and redshift. An important aspect of this study is that we used a consistent method for defining both the host galaxies stellar mass and the environment of AGN since there is wide range in how "environmental density" is actually quantified in the literature \citep[see e.g.][and references therein]{Krefting2020}. 

However, our paper and most existing studies rely on photometric redshifts for the characterization of the physical properties of the AGN and their hosts as well as deriving galaxy densities. This means significant uncertainties on all of these derived properties. However, upcoming large spectroscopic surveys out to cosmic noon \citep[e.g. the Prime Focus Spectrograph galaxy evolution survey][]{Greene2024} will be able to address more definitively the question of how AGN triggering depends on environment across different cosmic epochs. In particular, they will be able to verify whether our conclusion of a reversal in the density dependence of AGN at $z\sim1.2$ is correct or not. 

\subsection{Implications for AGN triggering and quenching}

The question of how AGN triggering and quenching depend on environment is closely related to the question of how galaxies star-formation and quenching depend on environment since both fundamentally are tied to the availability of gas \citep[see e.g.][]{Alberts2022a,Alberts2022b}. The environment can affect the supply of gas differently at different cosmic epochs and mass scales. For example, the environment can negatively impact the supply of gas due to processes such as ram-pressure stripping, tidal stripping and strangulation. This is especially impactful in lower mass galaxies leading to environmental quenching especially below $z\sim1$ \citep[e.g.][]{Peng2010}, although this may extend to higher redshifts \citep[e.g.][]{Taylor2023}. 

Conversely for higher mass galaxies, where AGN predominantly live, a denser environment can also enhance the supply of gas especially at cosmic noon through increased cold gas accretion as well as increased merger rates. For example, \citet{Waterval2025} find that for massive halos ($M_{halo}>10^{12}$M$_{\odot}$) we expect filaments of cold gas to flow down to the central galaxies at cosmic noon. This cold gas, coupled with enhanced rates of gas-rich mergers, fuels both star-formation and AGN activity in these halos. Indeed some studies do find enhanced star-formation rates in higher density environments at cosmic nooon, in contrast to the trends below $z\sim 1$ \citep[e.g.][]{Taamoli23}. The finding of \citet{Powell2022} that black hole mass depends both on the stellar mass of the galaxy and the mass of the host dark matter halo is in line with higher density environments also fueling greater accretion onto the black holes. However, this process largely stops below $z\sim 1-1.5$ due to shock-heating of the accreting gas in high mass halos (``hot-mode accretion") \citep[e.g.][]{DekelBirnboim2006,Waterval2025}. 

Our finding that, at $z>1.2$, at a given stellar mass, galaxies in denser environments are more likely to host IR AGN whereas at $z<1.2$ they are less likely to host IR AGN is consistent with the idea that in the higher mass halos, the supply of cold gas to the central galaxies, through accretion or mergers, is shut off at lower redshifts.  Below $z\sim1.2$ we find that AGN activity, at least the high accretion rate one we probe with our IR samples, shifts towards less dense environments (groups or field galaxies). This is in line with \citet{Kauffmann2004} who argue that for the nearby Universe, secular processes rather than mergers dominate AGN triggering. This finding is also in line with the high density environments at $z<1$ being dominated by red and dead galaxies \citep[e.g.][]{Peng2010}. 

The non-detection of these processes affecting the occurrence of unobscured X-ray detected AGN (both seen here and in \citet{Yang2018}) 
may be interpreted as 
the triggering of unobscured AGN associated with young star forming galaxies are more tied to local environments of their host galaxies rather than the conditions of the large scale extragalactic environments. However, we caution that X-ray and IR-selected AGN have different relative selection biases with respect to ther host galaxy properties, including stellar mass, dustiness and star-formation rate \citep[e.g.][]{Azadi2017,Coleman2022,Ji2022}. While we control here for stellar mass we do not control for any other host galaxy property especially star-formation rate.  
For example, \citet{Mountrichas2023} find that $log(L_X/[erg/s])>43$ AGN more frequently show evidence of a recent burst of star-formation and are less likely than the non-AGN to be bulge-dominated in higher density environments. They argue that this points to a common mechanism, such as mergers, triggers both the star-formation and the AGN activity in these higher density environments. 
Lastly, the presence of an AGN is expected to also contribute to the quenching of galaxies at higher masses. Therefore going beyond the current star-formation rate to study the star-formation histories vs. AGN incidence in sources in different density environments should elucidate this connection as well \citep[e.g.][]{Mao2022}. 

\section{Summary \& Conclusions} \label{sec:summary}

In this paper we study the effect of local galaxy density on the incidence of AGN while controlling for stellar mass. To test for redshift effects, we perform our analysis in two redshift bins, $0.1<z<1.2$ and $1.2<z<1.6$. To test for any potential differences with AGN selection, we also performed our analysis using X-ray and IR selected AGN as well as the a subset of heavily obscured sources that are IR-selected but not X-ray detected. We further verify our conclusions using a bootstrapping analysis. Our findings are as follows:

\begin{enumerate}
     \item Consistent with earlier studies, we find no environmental dependence for X-ray AGN when controlling for stellar mass. 
     \item By contrast, for both IR samples we find an environmental dependence (at $\approx$2\,$\sigma$ significance) in their likelihood to host an AGN among the highest stellar mass galaxies (i.e. those with stellar mass of $>2\times10^{11}$M$_{\odot}$ assuming Salpeter IMF). 
     \item The environmental effects are redshift dependent: at $z>1.2$, galaxies in denser environments are more likely to host IR AGN; but at $z<1.2$, galaxies in less dense environments are more likely to host IR AGN. 

\end{enumerate}

\begin{acknowledgements}
We are grateful to the anonymous referee for their careful reading and feedback which significantly improved the clarity of our paper.
We thank Jack Runburg for providing the initial IR and X-ray AGN catalogs for the XMM-LSS field. JR would also like to thank Ryan Hickox for useful discussions. This work is based in part on observations made with the Spitzer Space Telescope, which was operated by the Jet Propulsion Laboratory, California Institute of Technology under a contract with NASA. Based on observations obtained with XMM-Newton, an ESA science mission with instruments and contributions directly funded by ESA Member States and NASA. This work received support from the Massachusetts Space Grant Consortium (MASGC), one of 52 Space Consortia, established by NASA under the National Space Grant College and Fellowship Program created by Congress in 1987.   
\end{acknowledgements}

\software{Astropy \citep{astropy:2013, astropy:2018, astropy:2022}, TOPCAT \citep{topcat2011}}

\bibliography{agn_environment}{}
\bibliographystyle{aasjournalv7}

\end{document}